# An Empirical Study of Spam and Prevention Mechanisms in Online Video Chat Services

Xinyu Xing[1], Junho Ahn[2], Wenke Lee[1], Richard Han[2], and Shivakant Mishra[2]

[1] Georgia Institute of Technology
{xinyu.xing, wenke}@cc.gatech.edu
[2] University of Colorado - Boulder
{junho.ahn, rhan, mishras}@cs.colorado.edu

**Abstract.** Recently, online video chat services are becoming increasingly popular. While experiencing tremendous growth, online video chat services have also become yet another spamming target. Unlike spam propagated via traditional medium like emails and social networks, we find that spam propagated via online video chat services is able to draw much larger attention from the users. We have conducted several experiments to investigate spam propagation on Chatroulette - the largest online video chat website. We have found that the largest spam campaign on online video chat websites is dating scams. Our study indicates that spam carrying dating or pharmacy scams have much higher clickthrough rates than email spam carrying the same content. In particular, dating scams reach a clickthrough rate of 14.97%. We also examined and analysed spam prevention mechanisms that online video chat websites have designed and implemented. Our study indicates that the prevention mechanisms either harm legitimate user experience or can be easily bypassed.

## 1 Introduction

Online video chat services such as Chatroulette [5], Omegle [17], myYearbook Live [16], Tinychat [21] and vChatter [22] have been gaining popularity over the recent years. Though video chat services are still new, the number of their users has significantly increased (over 500% since 2010). The world's largest online video chat system - Chatroulette - claims that there are 20 million unique visitors per month and approximately 35,000 online users at any given time [20].

A common feature of such online video chat systems is to pair strangers from around the world for webcam based conversations. During a conversation, a participant clicks a "Next" button to move on to a new partner. This behavior is also described as "spin". Online video chat websites use Adobe Flash 10.04 to display video and access the users' webcam. Adobe Flash's peer-to-peer network capabilities via RTMFP allow all video, audio and text streams to be exchanged directly between user computers, without involving the server and hence without using any server bandwidth.

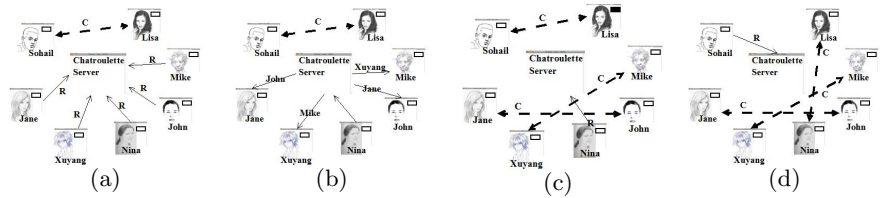

**Fig. 1.** A typical online video chat service.

A key problem in these video chat websites is that a large number of spam bots are abusing them to propagate spam. Generally, spammers execute voluminous copies of spam bots on many computers, and propagate unauthorized advertising information or phishing scams. These spam bots drive away legitimate advertising business, thus limiting such websites from obtaining more economic benefits. Furthermore, allowing spam bots to propagate scams on such websites exposes video chat users to malicious threats. To eliminate these spam bots, online video chat websites have developed and deployed their own prevention mechanisms. Unfortunately, our experience on these video chat websites indicated that the chances of encountering spam is still high. Take Chatroulette.com as an example. Every 10~20 spins include one spam. To understand the reason why such spam is persistent on online video chat websites, we investigated the characteristics of spam bots, revealed spamming workflow and examined video chat websites' prevention mechanisms. Exploring these is essential for helping online video chat providers enhance their prevention mechanisms and subsequently eliminate spam propagation. This paper provides an in-depth, empirical study of spam in online, video chat services. Based on this study, the paper suggests some prevention mechanisms.

Since online video chat systems employ a peer-to-peer architecture, monitoring spam, and investigating their behavior and characteristics is an extremely difficult task. Due to the absence of any centralized sever through which audio and video stream passes, it is not possible to put any monitoring probe. In this scenario, the only feasible solution is to collect data from a set of users who are participating in these services. We have employed this solution by building a spam bot program, registering several domain names, advertising these domain names via several spam bots, and recording the number of visits. During the experiments, our spam bots encountered chatters as well as other spam bots that spammers created, and recorded corresponding conversation sessions. Using these recorded session information along with Chatroulette web server logs, we conducted several measures and analysis of spam. In addition, our own spam bots are used for testing these spam prevention mechanisms of online video chat websites. Our study found out the largest spam campaign in online video chat websites is the spam carrying dating scams, which receives 14.97% of clickthrough rate. In contrast, spam carrying pharmacy scams receive less attention with at most 1.99% of clickthrough rate. Compared with pharmacy scams propagating via email, spam carrying pharmacy scams propagating via online video chat websites however is much more successful.

Using our deployment of spam bots in a small scale, not only do we witness the success of video spam at enticing tens of thousands of users, but also we examine online video chat websites' spam prevention mechanisms. Suprisingly, our spam bots can easily bypass the prevention mechanisms that Chatroulette and myYearbook designed and implemented. Though Omegle's spam prevention mechanisms block our spam bots, we observe that their spam prevention mechanisms also significantly harm user experience.

We summarize the contributions of this paper as follows.

- We have conducted several experiments on large scale and analysed the clickthrough rate of spam on the world largest online video chat website - Chatroulette.
- We present the first in-depth look at spam on Chatroulette, finding that the largest spam campaign is those carrying dating scams.
- We examine spam prevention mechanisms on the top three online video chat websites, exposing the security flaws of these prevention mechanisms.

We organize the remainder of the paper as follows. Secton 2 presents a brief background on video chat wesbistes. Section 3 analyses the clickthrough rate of spam and spam campaigns on a video chat website. Section 4 discusses the prevention mechanisms that online video chat websites designed and implemented. Section 5 discuesses some spam prevention suggestions, followed by conclusions in Section 6

## 2 Background

Figure 1 illustrates how an online video chat service such as Chatroulette, Omegle, or myYearbook works. A user first directs his/her web browser to the URL of the video chat's website to download its webpage (containing JavaScript, HTML and Adobe Flash files). The webpage requests the connection with another user from the video chat server, e.g. users Jane, Xuyang, Nina, John and Mike are making this request in Figure 1 (a). The server then returns the (IP) address of the computer host of another online user, in order to connect the two users together. For example, the server pairs Jane and John together, and Xuyang and Mike together in Figure 1(b). In most online video chat systems, pairing of users is conducted at random among all the users who are online at the same time. Therefore, the users do not know a priori the other person that they will be connected with for video conversations. Once a pairing is done, Adobe Flash code embedded within the page establishes a direct video session (a peer-to-peer connection) with the computer host of the other user, sending text, local camera and microphone data to the remote user while receiving that remote user's text, video and audio data. This is shown in Figure 1(c), where Jane and John have a video chat session established and Xuyang and Mike have a video chat session established. When a user clicks the "Next" button (user Lisa in Figure 1(c)), the browser requests a new user and his/her computer host's address, thus establishing a new video chat session (Lisa and Nina in Figure 1(d)).

Note that the server is chiefly involved in pairing users together by supplying their computer host addresses, and is not involved in relaying any video data.

### 2.1 Predicting Conversation Durations

When a user or his / her chatting partner clicks the "next" button, the user's web browser sends an HTTP post request to an online video chat server. By maintaining the web server's logs, an online video chat website can infer a user's conversation duration by analysing the timestamps of consecutive HTTP requests. Take the example of the user Jane shown in Figure 1 where she sends an HTTP request at 11:07:00 PM and starts a conversation with user John. After 45 seconds, John clicks "next" button, Jane's web browser automatically sends an HTTP request to Chatroulette server. Chatroulette server receives Jane's HTTP request at 11:07:45 PM and thus infers that the duration of Jane's conversation with John was 45 seconds.

### 2.2 Balancing Users' Requests

For some online video chat websites such as Chatroulette and Omegle, there are a large number of online users at any given time. For example, Chatroulette has 20,000 to 30,000 users at any given time, which is similar to Omegle. To balance such users' requests, online video chat websites distribute users and their HTTP requests to several servers. For example, Chatroulette used 4 to 6 web servers to process their users' requests. Note that the number of web servers relies on the number of online users. Since traffic balance is based on the granularity of users, users who connect to one Chatroulette server cannot have a conversation with others who connect to other Chatroulette servers. On average, there are 4,000 to 5,000 users on each Chatroulette server. Since online video chat users are randomly distributed across several web servers, the population of the users on each Chatroulette server in terms of age, gender and location remains approximately consistent.

### 2.3 Obtaining Snapshots

In online video chat services, users' text, video and audio data are transmitted via Adobe Flash's peer-to-peer networks. A few online video chat websites use HTTP pulling techniques to peroidically collect their users' snapshots. Using HTTP pulling techniques helps online video chat websites censor inappropriate video content (e.g., pornographic content and unauthorized advertisement) [4].

## 3 Spam on Chatroulette

Based on our extensive observation on online video chat websites, we noticed spammers typically use two approaches to perform spam propagation. One, a spammer uses a virtual webcam software to play an animated GIF image where a corresponding spam URL is explicitly displayed. Another approach is that a spammer replays a pre-recorded phishing video where an attractive woman sits in front of a webcam and types on her keyboard. Along with the phishing video, the spammer also runs a chatbot program that allows a spammer's spam bot to have a capacity to text chat with his chatting partner. A victim may mistakenly think that he/she is having a conversation with a real human chatting partner. Therefore, when a chat bot program sends the victim a spam URL, the victim may trust the chat bot and thus visit the spam URL. In this paper, we describe these two spamming approaches as image spam bot and video spam bot approach, respectively.

### 3.1 Methodology

There is no way to fully monitor spam bots and measure spam clickthrough rates because of Adobe Flash's peer-to-peer networks. To investigate spam in the context of online video chat websites, we registered several domain names, built our own spam bots and engaged in spamming activities. To avoid negative impact on online video chat services associated with our spam experiments, we displayed an explicit notification to those visitors who our spam bots coerced into our spam experiment websites. In our notification page, we described our experiments and re-directed visitors back to online video chat websites in 5 seconds. Our spam experiments were conducted on the largest online video chat website - Chatroulette due to the following considerations. First, the Chatroulette development team provides us with an internal data traces. We can use these data traces along with our spam bot data traces to conduct our spam analysis. Second, Alexa.com [1] reports show that online video chat websites such as Chatroulette, Omegle, TinyChat, myYearbook have similar populations in terms of visitors' age, ethnicity, education levels, and gender. Because these are major factors that affect the clickthrough rate of a spam URL, our experiments on Chatroulette are representative for other online video chat websites.

We built our spam bots with four major functions. (1) Our spam bot has the capacity of chatting with other Chatroulette users and playing a pre-recorded video. (2) Our spam bot has the capacity of taking snapshots of Chatroulette users while connecting to other Chatroulette users. (3) Our spam bot has the capacity of logging conversations. (4) Our spam bot can also log other Chatroulette users' IP addresses. All of these logs along with Chatroulette internal data traces are used for our spam study.

### 3.2 Clickthrough

Our prior experiences with Chatroulette and Omegle indicate that the most commonly seen spam was the spam carrying online dating scams. Therefore, our first experiment focused on measuring the clickthrough rate of online dating spam. Though the chance of encountering pharmacy spam is much lower than that of seeing dating spam on Chatroulette and Omegle, we still designed our second experiment to measure the clickthrough rate of pharmacy spam on Chatroulette.

The reason to do this is to investigate whether an online video chat website is a highly successful platform for coercing users to visit a spam URL. We compared the results of our pharmacy spam experiments with the results presented in [12] where pharmacy spam is sent through emails.

**Table 1.** Spam experiments and spam bots

| Group | Spam bot | Spam | unique chatter # | unique visitor # | Avg. conversation # per min. |
|---|---|---|---|---|---|
| 1 | image | dating | 811,317 | 98,113 | 3.76 per spam bot |
| 2 | image | pharmacy | 1,671,840 | 12,960 | 7.74 per spam bot |
| 3 | video | dating | 538,559 | 80,639 | 2.49 per spam bot |
| 4 | video | pharmacy | 985,091 | 19,636 | 4.56 perspam bot |

We executed 20 copies of our spam bot program on 20 Amazon's EC2 micro instances with Windows installations for one month period. We divided these 20 spam bots into 4 groups, and each group has 5 spam bots running the same spam experiment. To receive different audiences for different spam experiments, each of the four groups performs a corresponding spam experiment on a different Chatroulette server. Table 1 describes four spam experiments that are assigned to four groups and the corresponding spam bots that four groups perform. Similar to other video spam bots on Chatroulette, our video spam bots use a pre-recoded video with a fully-clothed young woman sitting in front a webcam and typing on her keyboard. Our video spam bots carrying dating and pharmacy spam used the same chatting scripts as those that other Chatroulette spammers used. Similarly, our image spam bots carrying dating spam used a similar image that other Chatroulette spammers used but with a different spam URL on it. As we did not find any spammers use image spam bots to send pharmacy spam, here we used an image that the notorious spam brand - canadian pharmacy [3] - used and edit it with our spam URL.

Table 1 shows the results of our spam experiments over a one-month period. Surprisingly, dating spam obtains tremendous audiences and high clickthrough rates (12.09% as well as 14.97% for video and image spam bot, respectively). These high clickthrough rates of dating spam imply that Chatroulette users engaged in their service with the purpose of looking for either virtual friendship or online dating opportunities. Unsurprisingly, notorious pharmacy spam has lower clickthrough rates than dating spam. However, this was by no means unsuccessful. Comparing with prior pharmacy spam carried via emails with a clickthrough rate of 0.00303% [12], pharmacy spam carried by video spam bots on Chatroulette had an approximately 650 times higher clickthrough rate than pharmacy spam on emails (0.00303% vs. 1.99%). In addition to these high clickthrough rates, our another interesting observation is that video spam bots typically achieved slightly higher clickthrough rates than image spam bots, especially when carrying the same spam. The reason behind this may be because of the characteristics of video spam bots. Since a video spam bot has the function of a chat bot, it has a higher possibility of persuading a victim to visit a spam URL by using its pre-defined chatting script.

Different from email spam that can be efficiently delivered to end users in a large scale, the breadth of spam delivery is dependent upon the number of chatters that a spam bot encounters. However, the number of chatters that a spam bot can encounter is not the only factor that determines how many visitors a spam URL can obtain. As shown in Table 1, the clickthrough rate of a spam URL is highly correlated to the average number of conversations per minute. That is to say, the longer a spam bot can have a conversation with a chatter, the more likely that a spam URL would be visited. It is quite obvious that a high clickthrough rate is not the goal that a spammer expect to achieve. Rather they strive to trick as many visitors as possible to visit their spam URL within the limited time. As a result, the most efficient spam bot program for advertising online dating spam are those image spam bots. Our prior observations echoed with this findings, i.e., image spam bots carrying online dating spam are more commonly used than video spam bots.

### 3.3 Spam and Affiliate Program

Using Chatroulette data traces along with the log data that our spam bots collected during our spam experiments, we first performed statistical analysis for those spam on Chatroulette. Chatroulette data traces contain 32,441 online users' snapshots and their corresponding IP addresses, covering all online users from 4:52:33 PM to 5:23:42 PM on 29 September 2011. We selected all snapshots which include explicit spam URLs. We labeled those IP addresses associated with selected snapshots as image spam bots' IP addresses. Since video spam bots used text messages to carry spam URLs via Adobe Flash's peer-to-peer networks, users' snapshots and corresponding IP addresses of Chatroulette data traces cannot help us identify those video spam bots on Chatroulette. To address this problem, we searched for those IP addresses that appear in Chatroulette data traces from our spam bots' logs. If the IP addresses also appear in our spam bots' logs, we extracted the corresponding text conversations and examine whether URLs are included in the text conversations. If an URL was detected, we labeled this corresponding IP address as an IP address which hosts video spam bots. Though the overlapping IP addresses between Chatroulette data traces and spam bots' logs are just a few, this could still help us identify most of video spam bots because spammers typically run their spam bots 24 by 7.

Table 2. Spam categories on Chatroulette.

| Category | Spam URL # | IP address # |
|---|---|---|
| Dating | 38 | 200 |
| Pharmacy | 4 | 25 |
| Malware | 3 | 11 |
| Make money fast | 2 | 6 |
| Hiring webcam models | 1 | 2 |

Aggregating all IP addresses that host either video spam bots or image spam bots, we obtained a list of 244 unique IP addresses advertising 48 unique spam URLs. Table 2 categorizes these 48 spam URLs into five categories. Unsurprisingly, dating spam was the largest spam campaign on Chatroulette and is advertised by 200 unique IP addresses. The number of pharmacy spam, to our surprise, was much lower on Chatroulette though our experiments show it can reach higher clickthrough rates ( see Table 1). To explain this, we revisited Table 1 and observed that the 5 video spam bots carrying pharmacy spam need a month to encounter approximately 1 million chatters and only coerced approximately 20 thousand chatters into visiting a spam URL. Apparently, this is not as efficient as what email spam does, which can send pharmacy spam to 350 million audiences at one time and obtain approximately 10 thousand visitors in a couple of days [12]. We further investigated the hosts of the spam bots found on Chatroulette by looking up the 244 spam IP addresses. To our surprise, we found that 224 out 244 IP addresses come from commercial cloud servers including Amazon.com [2], Rackspace.com [18], Slicehost.com [19] and Godaddy.com [11]. Different from those email spammers who use the power of botnets to advertise their spam URLs, spammers on Chatroulette use commercial computation resources to run their business.

**Table 3.** Spam URLs belonging to affiliate program streamate.com.

| Spam URL | IP address | Checkout URL |
| --- | --- | --- |
| randomdatingservice.com | 69.90.89.126 | https://secure.streamate.com/signup/?AFNO=1-0-627111-348978... |
| chatroulettehof.com | 69.90.89.126 | https://secure.streamate.com/signup/?AFNO=1-0-609824-344299... |
| omeglegirls.com | 69.90.89.126 | https://secure.streamate.com/login.php?AFNO=1-609824-JD-3-1-300... |
| omeglevideos.com | 69.90.89.126 | https://secure.streamate.com/login.php?AFNO=1-627111-JD-3-1-300... |
| xxxzap.com | 184.168.174.1 | https://secure.streamate.com/signup/?xsid=XGC&AFNO=1-623506-JD-3-6-355... |
| babezap.com | 184.168.63.1 | https://secure.streamate.com/signup/?xsid=XGC&AFNO=1-625703-JD-3-6-355... |
| scenecams.com | 46.252.206.1 | https://secure.streamate.com/signup/?xsid=XGC&AFNO=1-618951-JD-3-6-355... |
| chatraw.com | 50.63.36.1 | https://secure.streamate.com/signup/?xsid=XGC&AFNO=1-629669-JD-3-6-355... |

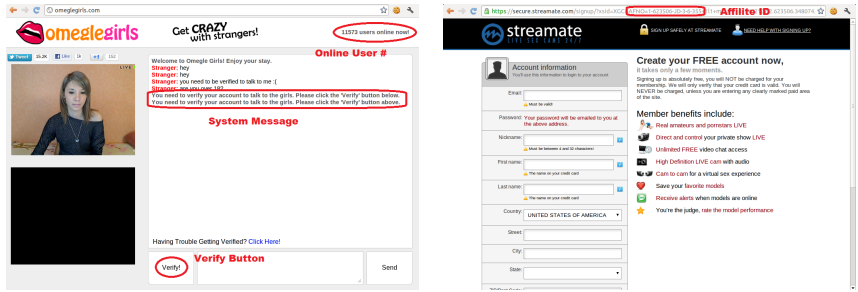

(a) A spam website template.   (b) A check out page of streamate.com.

**Fig. 2.** A website that a spam URL links to.

Since 80% of spam URLs include online dating services, we further investigated these online dating websites by visiting all 38 dating spam URLs. The purpose of our investigation was to see whether those online dating spam are unauthorzied advertisement or online dating scams. Taking the affiliate program of streamate.com as an example, we observed 8 out of 38 spam URLs including online dating services belong to this affiliate program. Table 3 shows these 8 spam URLs. To dissect this affiliate program, we first revealed the spam workflow. Similar to pharmacy spam introduced in [14], spammers first use streamate.com's advertising templates to create their own websites. They then registered several domain names, associate these domain names with their websites and use spam bots to advertise their domain names on online video chat websites. When a victim visits a spammer's website, he/she sees a page that presents a few things which appear to be real, but actually are not. Figure 2(a) shows an practical example. First, an online dating users' number appears on the right corner of the page. However, after examining the page's source codes, we easily found that the number is generated by a JavaScript function below rather than actual online users.

```
var usersonline = function(){
  var randomnumber=Math.floor(Math.random()*101)
  var time=Math.floor(Math.random()*2101)
  totalusers = totalusers + randomnumber
  document.getElementById('usercount').innerHTML = totalusers;
  setTimeout('usersonline();', time);
}
```

Second, there comes what appears to be a live chat window, which turns out to be a pre-recorded 20-second video of a chatter pretending to engage in a conversation with the victim. If the victim attempts to type into the fake chat field, the page refreshed with a totally different video of a totally different chatter and a system message - "Your Message was not sent! To Chat you must click 'VERIFY' to verify your age." - will post on the chatting text window. If the victim clicks "VERIFY" button, he/she will be redirected to a check out page provided by affiliate program streamate.com. Affiliate program streamate.com uses affiliate IDs to identify their affiliates. Therefore, when a victim is redirected to the check out page of streamate.com, a corresponding affiliate ID is also passed to streamate.com. As shown in Figure 2(b), this affiliate ID usually appears on the victim's address bar. Table 3 reveals the affiliate ID that each spam URL is associated with (i.e., AFNO). As some spammers engage in multiple affiliates to increase their economic benefits, an interesting observation in Table 3 is that the first four spam URLs are associated with the same IP address.

In addition to those spam websites, we further conduct a meta analysis of streamate.com. To determine whether streamate.com is a reliable dating service provider, we searched the Internet and find out several evidences concerning credit card frauds [9][6][15] (to name a few), indicating that affiliate program streamate.com is an online dating scam.

### 3.4 Spam Bot Prediction

Comparing with 32,441 online users, 244 users are associated with spam bots, which seems trivial. However, our experiences on Chatroulette and Omegle indicate that a chatter typically encounters a spam bot for every 10∼20 spins on average. It is apparent that spammers execute multiple copies of their spam bot programs on each machine. To predict the average number of spam bot programs running on each machine, we analyse Chatroulette's web server log. Taking a one-hour Chatroulette web server log as an example, we observe that one Chatroulette server receives 350 HTTP requests in every second on average (see Figure 3). Using this log information, we predict the number of copies of spam bot programs running on Chatroulette. Let $n$ be the number of spam bot programs which send HTTP request to Chatroulette server at any given time. Then the probability that a user encounters a spam bot when he/she sends HTTP request to Chatroulette server is $\frac{n}{350}$. The probability that a user encounters one spam bot at the $i$th spin ($p$) then can be described as following.

$$p = 1 - (1 - \frac{n}{350})^i \qquad (1)$$

Assuming $n = 50$, we can calculate the probabilities that a user encounters a spam bot program at the 10th and the 20th spin are approximately 79% and 95%, respectively. Recall that Table 1 shows the average shortest conversation is 7.75 seconds per spam bot program (e.g., an image spam bot program carrying pharmacy scams can talk to 7.74 chatters in every minute), while the average longest conversation is 24.10 seconds per spam bot program (e.g., a video spam bot program carrying dating spam can talk to 2.49 chatters in every minute). Based on these information, we predict that the lower and the upper bound of the total number of spam bot programs running on one Chatroulette server are 388 and 1205. Note that this range 388~1,205 only describes the total number of spam bot program running on one Chatroulette server. Since there are 6 Chatroulette web servers in total, we estimated the total number of spam bot programs hosting on Chatroulette are in the range of 2,328~7,230. In other word, each spammer's machine runs 10~30 spam bot programs.

Though our prediction above can well explain why a user can easily encounter a spam bot in every 20 spins, our prediction is far exceeding our expectations. According to our observation and experience, we found that Chatroulette has implement and deployed an effective CAPTCHA mechanism that can detect those spammers who attempt to execute multiple spam bots on a single machine. To further verify the correctness of our prediction, we examine Chatroulette's spam prevention mechanism along with other spam prevention mechanisms that other online video chat websites designed and implemented in the following section.

## 4 Prevention Mechanisms

To curb the abuses of spam, some online video chat websites design and implement their own spam prevention mechanisms. To the best of our knowledge, such websites include Chatroulette, Omegle and myYearbook. However, our observations on these websites indicate that there are still a great number of spam on these websites. To find the reason for the abuses, we examine these prevention mechanisms that are implemented on these online video chat websites and analyse their security flaws. In addition, we also discuss the usability of state-of-the-art mechanisms for spam bot prevention.

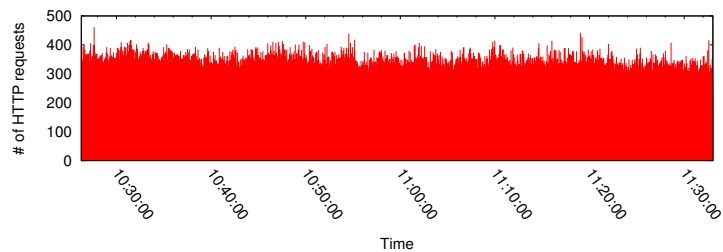

**Fig. 3.** HTTP requests on a Chatroulette web server.

### 4.1 CAPTCHA based Prevention

Both Chatroulette and Omegle design and implement their own CAPTCHA based prevention mechanisms to eliminate spam bots in their systems. The main idea of their prevention mechanisms is to analyse the characteristics of users' HTTP requests. When pre-defined spam characteristics are detected, CAPTCH challenges will be sent to corresponding users, i.e., possible spammers.

**Prevention Workflow** Omegle's CAPTCHA based spam prevention mechanism is an IP-based spam prevention approach. It uses a pre-defined users' HTTP request pattern to determine whether a user is a spammer. Their pre-defined pattern is based on two assumptions. (1) Every user connects to Omegle using an unique IP address. (2) If a spammer executes multiple copies of a spam bot program from a single machine, the frequency of the spammer's HTTP requests to Omegle servers is high. When Omegle servers receive HTTP requests from a same IP address at a high frequency, Omegle invokes Google RECAPTCH API sending a CAPTCHA challenge to the suspected IP address. Omegle does not respond to HTTP requests from the suspected IP address until the CAPTCHA challenge is correctly solved.

Similar to Omegle, Chatroulette's spam prevention mechanism is also based on a CAPTCHA technique. However, the pre-defined rule for triggering a CAPTCHA challenge is different. Chatroulette's prevention mechanism uses the number of active HTTP sessions to determine whether a user is a possible spammer. When a user connects to a Chatroulette web server for the first time, Chatroulette assigns a temporary session ID to the user. Therefore, when a spammer attempts to establish multiple connections to Chatroulette servers, Chatroulette observes multiple active sessions from a same IP address. Using this intuitive characteristic, Chatroulette determines whether a CAPTCHA challenge needs to be sent to an IP address. According to probes using our spam bots, we found out the maximum threshold of the number of active sessions for triggering a CAPTCHA challenge is equal to four. In other word, if there is an IP address associated with four active sessions, a Chatroulette's CAPTCHA challenge will be delivered to the IP address.

**Security Flaws** Using 10 copies of our spam bot to examine Omegle's spam prevention mechanism, we observe our machine that executes these spam bot programs receives a Google CAPTCHA and is temporarily blocked after our spam bots are launched for one minute. To further verify that Omegle's spam prevention is based on IP addresses, we delete all the Local Shared Objects (LSO) and find that our reconnection attempts failed. In addition, we also experiment with another machine behind the same NAT and receive the same connection failure. It is apparent that Omegle's spam prevention mechanism is sufficiently effective for keeping spammer's spam bots away from their system, especially when a spammer attempts to execute several spam bot programs on a single machine. However, a security flaw still exists when we experiment with a single spam bot program advertising a spam URL on Omegle. Except for an image spam bot carrying pharmacy spam that receives a Google CAPTCHA challenge after being launched for 17 minutes, we surprisingly observe that all four spamming combinations used in our Chatroulette's clickthrough experiments successfully bypass Omegle's spam prevention without experiencing Google CAPTCHA challenges in a three-day experiment. The reason of this success is straightforward. Omegle uses the frequency of user's HTTP requests to distinguish a spammer. Tabel 1 shows that all our spam bots except the image bot carrying pharmacy spam usually have longer conversations with users. As a result, Omegle cannot expect our spam bots' HTTP requests at a high frequency except the image spam bot carrying pharmacy spam.

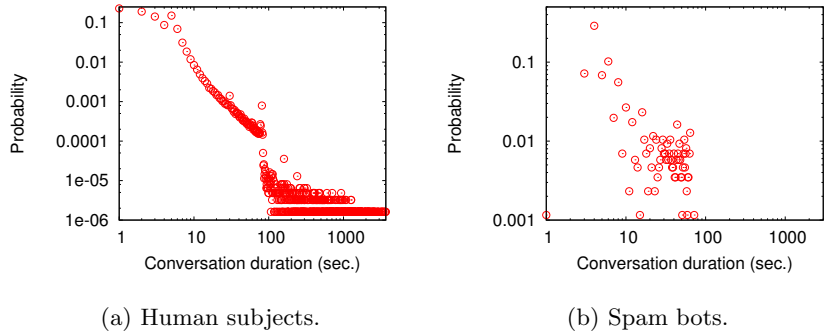

(a) Human subjects.     (b) Spam bots.

**Fig. 4.** Distributions of human subjects' and spam bots' conversation durations.

In addition to the security flaws above, we also examine whether Omegle's prevention mechanism may harm their users' experience if it is implemented by an online video chat website such as Chatroulette. We extensively involve in Chatroulette by having conversations with Chatroulette users. During a conversation, a subjective Turing test is involved, which asks our subjects to flash a peace sign or place their fingers upon their head. There are 5,702 subjects passing our Turing test in a one-month period. To support this study, Chatroulette provides us with a data trace associated with these 5,702 subjects which includes the conversation durations and snapshots of each subject. These subjects totally conducted 1,105,320 conversations in a one-month period. We compare the conversation characteristics of these human subjects with those of our spam bots. Figure 4 shows the probability distributions of human subjects' conversation durations and spam bots' conversation durations in a log-log scale. An interesting observation is that both human subjects and spam bots have more short conversations than long ones. If Omegle's prevention mechanism uses consecutive short conversations to distinguish a spammer, it is highly likely to mistakenly suspect a legitimate user to be a spammer because both human users and spammers have more short conversations.

Compared with Omegle's spam prevention mechanism, Chatroulette's spam prevention mechanism is easier to be bypassed. As Chatroulette uses the maximum threshold of the number of active sessions to trigger their CAPTCHA challenges, a spammer can bypass Chatroulette's spam prevention by sharing a same session ID with all copies of his spam bot program. To verify this, we execute 15 image spam bot programs carrying pharmacy spam on a single machine in a three-day period. Unsurprisingly, our three-day experiment receives no Chatroulette's CAPTCHA challenges.

### 4.2 Session based Prevention

Different from Chatroulette and Omegle, myYearbook design and implement a complex session based prevention mechanism to curb the abuses of spam. Similar to CAPTCHA based prevention, session based prevention cannot completely stop spamming activities. However, myYearbook's prevention mechanism may effectively avoid multiple spam bot programs running on a single machine. It is essential for reducing spamming activities because it increases both the difficulty and economic cost of spamming activities.

myYearbook's prevention mechanism involves a two-sided effort - a chatting client side and myYearbook's web server side. myYearbook's chatting client is implemented using Adobe Flash, JavaScript and HTML codes; myYearbook's server side is implemented using PHP. To dissect the session based prevention mechanism, we download Flash applications, JavaScript and partial HTML codes of myYearbook's chatting client. Since myYearbook has not obfuscated and encrypted their Flash applications, we use Flash Decompiler Trillix [8] to decompile myYearbook's Flash applications and analyse myYearbook's spam prevention workflow.

**Prevention Workflow** We first dissect the workflow of myYearbook by observing the HTTP traffic between a chatting client and myyearbook.com. Though myYearbook is a sign-in required onlive video chat service that requires each user register an account on their website and uses his/her account to access myYearbook service, all the HTTP traffic except for the traffic in the sign-in process between the chatting client and myYearbook.com is through an unencrypted channel (i.e., no HTTPS are applied). Figure 5 and 6 show the prevention workflow of myYearbook.

In Figure 5, a chatting client first sends an authentication request to myYearbook.com. If the authentication process passes, several Javascript, Flash and HTML files are downloaded to the chatting client. On the chatting client side, the downloaded Flash code - MessageBridge.swf - is performed to check whether the chatting client has already initiated a conversation session. Note that the term - *conversation session* - in this paper means the session established between a chatting client and myYearbook.com, but not the chatting session established between two chatting clients. If there is no conversation session on a chatting client, the Falsh code calls function setEnabledFlag("videoReady",!0) and never() of Javascript codes. The function setEnabledFlag("videoReady",!0) is invoked for establishing a conversation session on the chatting client, and function nerve() is invoked for initiating a heartbeat for the current conversation session and registers the conversation session on myYearbook.com. The heartbeat of a conversation session on the chatting client side is maintained by funtion nerve() and is periodically sent back to myYearbook.com for letting myYearbook.com know that the conversation session is active. In the current implementation, the heartbeat is sent back to myYearbook.com in every 5 seconds. On the server side, every time when an up-to-date heartbeat is received, the server (myYearbook.com) verifies the heartbeat. A conversation session is terminated if the server does not receive a heartbeat from the chatting client. In addition, the incorrect heartbeat sent from chatting client will be omitted. When the conversation session is active, the chatting client is able to send matching requests to myYearbook matching him/her with other chatting partners. The heartbeat of a conversation session is also associated with each matching request because the server needs to verify whether the matching request is legitimate or fraudulent. The server responds to a matching request by returning chatting partner's corresponding information including IP address and UDP port, etc.

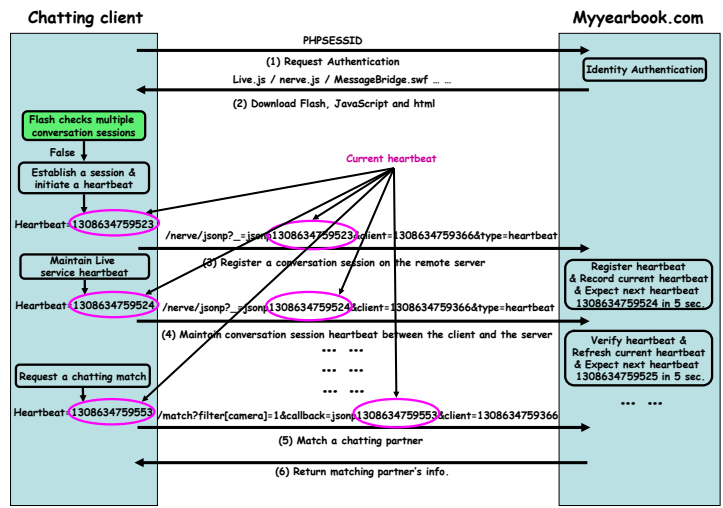

**Fig. 5.** myYearbook Live! workflow when only one conversation session is established.

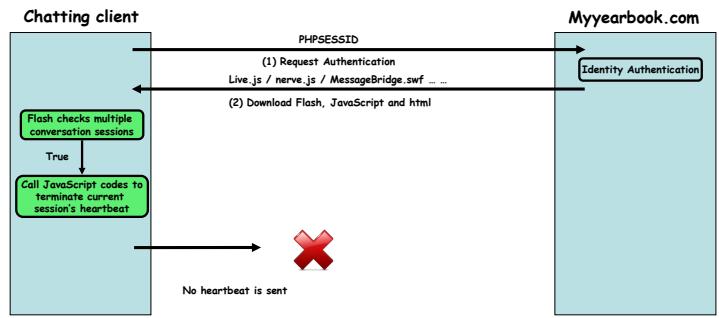

**Fig. 6.** myYearbook Live! workflow when attempting establish multiple conversation sessions.

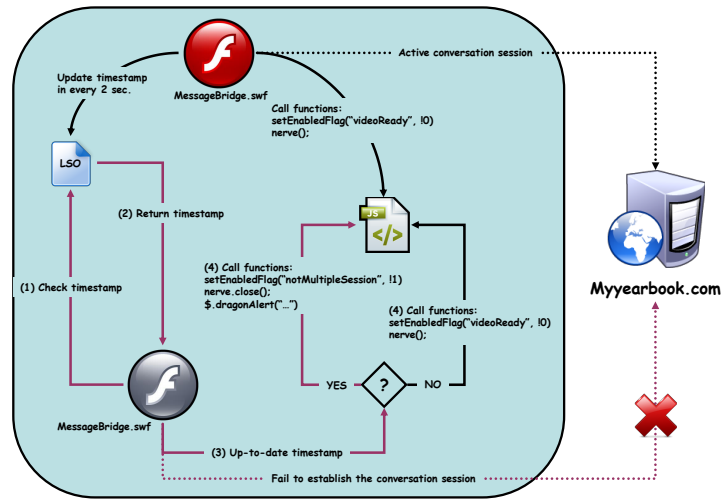

**Fig. 7.** Multiple conversation session detection workflow. The purple lines describe the entire detection process when a conversation session has existed.

On the contrary, if the downloaded Flash code - MessageBridge.swf - detects an existing conversation session on the chatting client, as shown in Figure 6, MessageBridge.swf invokes the JavaScript funtion - onMultipleSessions() that performs the following operations

```
a.behaviors.onMultipleSessions(
   function(){
      setEnabledFlag("notMultipleSessions",!1);
      nerve.close();
      $.dragonAlert("You can only chat ... ...")
   }
);
```

First, onMultipleSessions() invokes setEnabledFlag("notMultipleSessions", !1) that stops the chatting client establishing the second conversation session. It then invokes nerve.close() to terminate the initialization of a heartbeat for the second conversation session. Since myYearbook.com cannot receive a heartbeat from the second conversation session, it fails to register the second conversation session.

To determine whether a chatting client can establish a conversation session, myYearbook harnesses Adobe Flash codes implemented in file MessageBridge.swf. We decompile MessageBridge.swf and thus dissect the process of checking multiple conversation sessions. The major function of the Flash codes is described as follows.

```
1   public function MessageBridge() {
2       ... ...
3       this._sharedObjTimer = new
            flash.utils.Timer(SHARED_OBJ_UPDATE_INTERVAL);
4       this._heartbeatTimer = new flash.utils.Timer(60000);
5       ... ...
6       this.initSharedObj();
7   }
8   internal function initSharedObj():Boolean {
9       ... ...
10      this._sharedObj =
            flash.net.SharedObject.getLocal("myyearbook.com-MessageBridge",
            "/");
11      this._sharedObj.flush();
12      ... ...
13      if (this._sharedObj.data.name === this._dynamicName) {
14          this.startSharedObjTimer();
15          flash.external.ExternalInterface.call("Platform_messageBridge_
16          ready");
17      }
18      else {
19          this.checkSharedObj();
20      }
21      ... ...
22  }
23  internal function
        checkSharedObj(arg1:flash.events.TimerEvent=null):Boolean {
24      var loc1:*=new Date();
25      ... ...
26      if (loc1.getTime() - this._sharedObj.data.timestamp <
            MULTIPLE_SESSIONS_THRESHOLD) {
27          if (arg1 !== null) {
28              flash.external.ExternalInterface.call("Platform_messageBridge_
29              handleMultipleSessions");
30          }
31          else {
32              this._sharedObjTimer.delay = MULTIPLE_SESSIONS_THRESHOLD;
33              this._sharedObjTimer.addEventListener(flash.events.TimerEvent.
34              TIMER, this.checkSharedObj);
35              this._sharedObjTimer.start();
36          }
37          return false;
38      }
39      ... ...
40  }
```

MessageBridge() is the constructor of class MessageBridge. It is invoked when a web browser loads MessageBridge.swf. MessageBridge() first initiates a timer for a local shared object (LSO - a Flash cookie) and heartbeat. The constructor then invokes internal funtion initSharedObj() to initiate an LSO for a current conversation session. In the LSO, a timestamp is stored and updated at a certain update interval (SHARED_OBJ_UPDATE_INTERVAL) until the current conversation session is terminated (i.e., chatting client shuts down its web browser). To prevent multiple conversation sessions being established, funtion initSharedObj() also checks whether an LSO has already existed on a chatting client

(i.e., whether a local shared object file - myyearbook.com-MessageBridge.sol has already existed). If the local shared object file has already existed, function checkSharedObj() is invoked. Function checkSharedObj() examines whether the timestamp in the existing local shared object file is up-to-date (i.e., the timestamp in the existing local shared object file only expired for less than MULTIPLE_SESSIONS_THRESHOLD milliseconds). If the timestamp in the existing local shared object file is up-to-date, the existing conversation session will continue to be active, while the new initial conversation session will be terminated by calling external JavaScript codes (which described above). To illustrate this, Figure 7 shows the entire operation process of Adobe Flash codes.

**Security Flaws** In Figure 5, 6 and 7, several security flaws can be easily seen. First, we recall that the Adobe Flash codes in MessageBridge.swf are mainly used for detecting multiple conversation sessions on the chatting client. To initiate or terminate a conversation session on the chatting client, the Flash codes need to invoke external JavaScript codes. Since JavaScript source codes are both public readable and writable, the JavaScript codes that perform conversation session termination can be freely modified by a spammer. Therefore, even though Flash codes in MessageBridge.swf detect multiple conversation sessions on a chatting client, the spammer can substitute the conversation session termination operations with the conversation session initialization operations, and thus establish multiple conversation sessions between the chatting client and myYearbook.com.

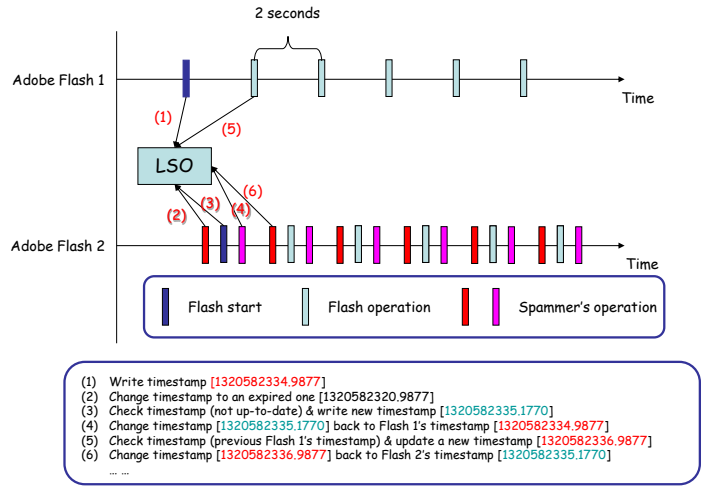

**Fig. 8.** Bypassing session based spam prevention by manipulating a LSO.

In addition to modifying JavaScript codes to bypass the session based spam prevention mechanism on a chatting client, another approach is to fool Adobe Flash operations. As shown in Figure 7, when a Flash program is executed in a web browser, it first checks the local shared object that is stored on the chatting client. If the timestamp stored in the local shared object file is up-to-date, this implies that there is another conversation session active on the chatting client. Current Adobe Flash implemention updates local shared object's timestamp at an interval of every 2 seconds. In other word, the time when a Flash program visits the LSO is predictable. To establish multiple conversation sessions on a single machine (i.e., spammers run multiple spam bot programs on a single machine), spammers therefore need to deceive Flash codes and make function checkSharedObj() believe the timestamp in the local shared object has expired for more than 2 seconds. To do this, we wrote an independent C program and used the program to control the timestamp of the local shared object - myyearbook.com-MessageBridge.sol. Figure 8 describes an example where a spammer fools two Adobe Flash programs by manipulating the timestamp stored in a local shared object. First, a spammer launches Flash program 1. When Flash program 1 starts, it writes its timestamp into a local shared object. Since Flash 1 visits the LSO in every two seconds, a spammer can predict Flash program 1's next visit to the LSO. Using this characteristic, the spammer modifies the timestamp stored in the LSO and changed the timestamp to an expired timestamp before he launches Flash program 2. Since Flash program 2 sees an expired timestamp in the LSO, Flash program 2 successfully starts and writes a new timestamp into the LSO. To ensure Adobe Flash 1 still works, the spammer also needs to change the timestamp in the LSO back to the timestamp that Flash program 1 stored before Flash program 1 revisits the LSO. Because Flash programs' visits to the LSO are predictable. Therefore, a spammer can easily bypass spam prevention mechanism. We verify the feasibility using our spam bot programs.

### 4.3 Discussion

Gianvecchio et al. [10] present a classification mechanism to distinguish chat bots from human in the context of the Internet chat room. The classification mechanism utilizes both chatting message size as well as inter-message delay characteristics to distinguish chat bots from human chatters. Though this classification mechanism shows reasonable good classification performance, online video chat websites still cannot obtain benefits from it because user's text messages, video and audio messages are all transmitted through Adobe Flash's peer-to-peer networks and there is no trival method to obtain users' chatting scripts from a centrial point. In addition, image spam bots do not use text messages to carry their spamming information, which may bypass the chat bot classification approach.

## 5 Suggestions and Impacts

Based on our examination of two major spam prevention mechanisms, we provide some spam prevention suggestions for online video chat websites. First, CAPTCHA based spam prevention is not suitable for online video chat websites that do not have a sign-in requirement such as Chatroulette, Omegle etc., because there is no effective method to identify a specific suspected user. Sending a CAPTCHA challenge to all the users behind the same NAT may harm legitimate users' experience. For those online video chat websites that require users' registration such as myYearbook and vChatter, CAPTCHA based spam prevention is a helpful mechanism if the scheme that triggers a CAPTCHA challenge has reasonable low chance to mistakenly treat a legitimate user as a spammer that runs tens of spam bots simultaneously.

Different from CAPTCHA based prevention, the session based spam prevention mechanism, though having several security flaws, can be useful if some improvements are added. The first suggestion for online video chat websites is to obfuscate and encrypt their Flash program using commercial Flash protection tools such as [7][13] etc., because a spammer may easily decompile their Flash program, understand their program logic, and even remove spam prevention functions from Flash program. Both Flash program obfuscation and encryption can make reverse engineering and the removal of prevention function extemely difficult. With the first assumption that our first suggestion implies (i.e., Flash program's obfuscation and encryption makes a Flash program unmodifiable), our second suggest is to eliminate side channel attacks by integrating JavaScript functions into a Flash program because JavaScript codes can be modified by spammers. Finally, the most important suggestion for session based spam prevention is to randomize the visit for a local shared object. Recall that a spammer can easily predict the time when Flash programs visit a local shared object. Using this knowledge, a spammer can manipulate the content of the local shared object, thus deceiving Flash programs to believe no other Flash programs are executing at the same time. By randomizing the Flash programs' visits to a local shared object and incorporating first two suggestions, any online video chat websites can sucessfully prevent multiple spam bot programs running on the same machine. Finally, we also note that both CAPTCHA based or session based spam prevention cannot completely eliminate the use of a single spam bot in online video chat websites. However, these prevention mechanisms are still useful for increasing spamming economic cost and reducing the number of spam bots.

Since our spam bot experiments indicate that a majority of spam bots are hosted on commercial cloud services, we report our study to both Chatroulette and Omegle. Based on our study and report, both video chat websites have already verified the spamming IP addresses that we report and blocked all the IP address ranges that belong to Amazon.com, Rackspace.com, Slicehost.com and Godaddy.com. The following is the comment from Leif K-Brooks - Omegle CEO. Leif said: *"That's surprising, and it's useful information. Thank you! It explains a lot about why spammers are so persistent..."*. Up to the time when we submit this paper, our one-hour consecutive observations from both Chatroulette and Omegle indicate that no spams have been found.

# 6 Conclusion

This paper conducts several spam experiments on Chatroulette. Using a real world data traces from Chatroulette along with our self-collected data traces, we investigate and analyse spam in the context of online video chat services. We find that spam on Chatroulette typically receives more attention than email spam. Furthermore, we predicte the lower and upper bound for the number of spam bots on Chatroulette. Motivated by our spam experiments, we also examine online video chat websites' spam prevention mechanisms. We find that most of prevention mechanisms can be easily bypassed. We reported our study to both Chatroulette and Omegle.

# References

1. Alexa web site. http://www.alexa.com/.
2. Amazon elastic compute cloud. `http://aws.amazon.com/ec2/`.
3. Spam trackers. `http://spamtrackers.eu/wiki/index.php/Canadian_Pharmacy`.
4. Nudity filter helps chatroulette clean up. January 20, 2011. `http://www.cbc.ca/news/technology/story/2011/01/20/algorithm-tech-nudity-chatroulette-filter-flashers.html`.
5. Chatroulette web site. http://www.chatroulette.com/.
6. Streamate.com complaints & reviews - stay away. April 18, 2011. `http://www.complaintsboard.com/complaints/streamatecom-c444119.html`.
7. Eramsoft software. http://www.eramsoft.com/.
8. Flash decompiler. `http://www.flash-decompiler.com/`.
9. Stupid but innocent, what do i do. December 13, 2008. `http://forum.freeadvice.com/arrests-searches-warrants-procedure-26/stupid-but-innocent-what-do-i-do-443421.html`.
10. S. Gianvecchio, M. Xie, Z. Wu, and H. Wang. Measurement and classification of humans and bots in internet chat. In *Proceedings of USENIX Security*, 2008.
11. Godaddy. `http://www.godaddy.com/`.
12. C. Kanich, C. Kreibich, K. Levchenko, B. Enright, G. M. Voelker, V. Paxson, and S. Savage. Spamalytics: An empirical analysis of spam marketing conversion. In *Proceedings of CCS, Alexandria, USA*, 2008.
13. Kindi software. http://www.kindi.com/.
14. K. Levchenko, A. Pitsillidis, N. Chachra, B. Enright, M. Felegyhazi, C. Grier, T. Halvorson, C. Kanich, C. Kreibich, H. Liu, D. McCoy, N. Weaver, V. Paxson, G. M. Voelker, and S. Savage. Click trajectories: End-to-end analysis of the spam value chain. In *Proceedings of IEEE Symposium on Security and Privacy*, 2011.
15. Streamate.com complaints — scams, frauds and reviews. November 05, 2011. `http://www.merareview.com/comments/3922`.
16. Myyearbook live web site. http://live.myyearbook.com/.
17. Omegle web site. http://www.omegle.com/.
18. Rackspace hosting. `http://www.rackspace.com/`.
19. Slicehost. `http://www.slicehost.com/`.
20. Chatroulette parts with private parts, looking for a new look. May 4, 2011. http://techcrunch.com/2011/05/04/chatroulette-parts-with-private-parts-looking-for-a-new-look/.
21. Tinychat web site. http://tinychat.com/.
22. vchatter web site. http://www.vchatter.com/.